\documentstyle[aps,prb]{revtex}

\begin{document}

\twocolumn[\hsize\textwidth\columnwidth\hsize\csname@twocolumnfalse\endcsname

\draft


\title{Density-matrix  renormalization-group
method in momentum space}

\author{T.  Xiang}

\address{ Research Center in  Superconductivity,  University
 of Cambridge,  Madingley  Road,  Cambridge  CB3 0HE, United
 Kingdom}

\date{\today}

\maketitle

\begin{abstract}

A momentum-space approach of 
the density-matrix renormalization-group
(DMRG) method is developed.  Ground state
energies of the Hubbard model are evaluated using this
method and compared with exact diagonalization as well as
quantum Monte-Carlo results.  It is shown that the
momentum-space DMRG is a very useful numerical tool for
studying the Hubbard model and other fundamental models of
interacting electrons in two dimensions.  For the Hubbard
model in two dimensions, the momentum-space DMRG method
reproduces accurately the exact diagonalization results of
ground state energies on a $4\times 4$ lattice and yields new
upper bounds of ground state energies on an 8$\times$8 lattice.

\end{abstract}


]

The real-space DMRG method, which was proposed by
White\cite{white} in 1992, is a powerful method for studying
ground state properties of interacting electrons or spins.
It has been successfully applied to various
one-dimensional (1D) quantum models, such as the spin
Heisenberg model\cite{white,heisenberg}, the Kondo lattice
model\cite{yu}, the Hubbard model\cite{noack}, and other
models\cite{other}.  However, in 2D its
application is still not satisfactory\cite{liang}.  In this paper we
generalize the DMRG method to momentum space.  We hope 
this may enlarge the range of the
application of the DMRG method and provide an accurate
numerical method for studying interacting electrons in 2D.
We shall take the Hubbard model as an example to show how
the DMRG method works in momentum space.  The properties of
the Hubbard model have been extensively studied in both one
and two dimensions, and so this provides a good background for
comparing the momentum-space DMRG method with 
other methods.

The DMRG method is a diagonalization technique which
attempts to use a small number of states, say $m$ states, to
expand the ground state (or some low energy excitation
states) accurately.  In conventional numerical
renormalization-group (RG) methods, one keeps the $m$ lowest
energy eigenstates of a block
Hamiltonian\cite{wilson,tao,white2}.  In the DMRG method,
however, one keeps the $m$ most probable basis states in
describing the ground state of a larger block, called a
superblock.  A superblock contains two blocks, a system
block and an environment block.  In a DMRG iteration, one
diagonalizes the Hamiltonian of the superblock, finds out
the reduced density matrix for the system block from the
ground state of the superblock, truncates the Hilbert
subspace by keeping the $m$ largest eigenstates of the
density matrix, adds one or more sites to the system block
to form a new superblock, and then repeats the above
procedure until the desired result is obtained.

There are two approaches in constructing a superblock, an
infinite-lattice approach and a finite-size approach.  In
the infinite-lattice approach, the environment block is
generally chosen as the space reflection of the system
block.  In the finite-size approach, on the other hand, the
size of the superblock is fixed and the environment block is
chosen as the remaining part of the lattice for a given
system block.  The infinite-lattice approach allows the size
of the superblock to be flexible and can be used to study
directly the thermodynamic limit.  The finite-size approach
is, however, more accurate in calculating
quantities for a finite lattice system.  In White's original
paper, two sites are added to a superblock each time, one to
the system block and the other to the environment block.
This way of constructing a superblock preserves the
reflection symmetry and avoids some problems caused by
odd-size lattices if the infinite-lattice approach is used
in 1D.  However, it is generally more efficient in
computation if one adds just one site to the system block
and no site to the environment block when the finite-size
approach is used.

The DMRG method satisfies the variational  principle because
it uses a small physical  subspace to  approximate  the full
Hilbert space and no  unphysical  states enter the truncated
Hilbert  subspace.  Thus the ground  state  energy  obtained
using  this  method  is  always a upper  bound  for the true
value.  The error for the ground state energy  decreases  as
the number of retained  states is increased.  The truncation
error  is  generally  smaller  than  the  true  error of the
result.

The   Hubbard   model   is   defined   by  the   Hamiltonian
\begin{equation} 
H=-t\sum_{<ij>\sigma}c_{i\sigma}^\dagger c_{j\sigma}
+U\sum_in_{i\uparrow}n_{i\downarrow}, \end{equation} where
$<ij>$ means summation over nearest neighbors.  In momentum
space, it reads \begin{equation} H=\sum_{k\sigma}\epsilon_k
c_{k\sigma}^\dagger c_{k\sigma} + {U\over
N}\sum_{k_1,k_2,k_3} c_{k_1\uparrow}^\dagger c_{k_2\uparrow}
c^\dagger_{k_3\downarrow} c_{k_1-k_2+k_3\downarrow}
\label{hamiltonian2}
\end{equation} 
with $N$ the lattice size.  Here the periodic boundary
condition is assumed.  Each momentum $k$ point has four
degrees of freedom, i.e.  $\{$ $|0>$, $c_{k\uparrow}^\dagger
|0>$, $c_{k\downarrow}^\dagger |0>$, $c_{k\uparrow}^\dagger
c_{k\downarrow}^\dagger |0>$ $\}$.  In the RG iteration,
generally a $k$ point with these fours states can be added
to a system block to form a new system block.  
However, in our calculation we treat the spin
degree of freedom as an extra spatial coordinate and a
momentum-spin point $(k\sigma )$, which has only two degrees
of freedom $\{ |0>,c_{k\sigma}^\dagger |0>\}$, as a basic
unit which is added to a system block.  In this case, the spin
rotation symmetry is broken, but the total number of degrees
of freedom of a superblock is reduced and more states can be
retained at each truncation of Hilbert subspace.  This can
reduce truncation errors and save computer time.

For the convenience in the discussion below, we label a
superblock as ($A$ $\bullet$ $B$), where $\bullet$
represents a momentum-spin point ($k\sigma$) which has just
been added to the system block, $A$ represents the collection of
all the momentum-spin points in the system block excluding
the point ($k\sigma$), and $B$ represents the collection of
all the momentum-spin points in the environment block.
There are many ways in ordering $k$ points in a system.  
In calculation, one should test a few possibilities of orders of
$k$ points and choose the one which gives the lowest ground
state energy.

The Hubbard interaction is local in real space.  However, it
is non-local in momentum space; it contains terms which link
two or more $k$ points in momentum space.  The summation in
the second term of (\ref{hamiltonian2}) contains $N^3$
terms.  They can be grouped into the terms which are defined
purely in each sub-block and the terms containing
interactions among $A$, $\bullet$, and $B$.  To evaluate
matrix elements of the Hamiltonian, we find it is very
useful to define the following composite operators of
electrons:  in block A,
\begin{eqnarray} 
{a}_0(p\sigma ) &=& c_{p\sigma } \delta_{(p\sigma ) \in A},
\nonumber\\
{a}_1(p\sigma )&=& \sum_{q}{a}^{\dagger}_0 ({q\sigma})
{a}_0({p+q\sigma}), \nonumber \\
{a}_2(p)&=&\sum_{q}{a}^{\dagger}_0({q\uparrow})
{a}_0({p+q\downarrow}), \label{composite} \\
{a}_3(p\sigma )&=&\sum_{q_1q_2} {a}^{\dagger}_0 ({q_1\bar
\sigma}) {a}_0 ({q_2\bar\sigma}) {a}_0 ({p+q_1-q_2\sigma})
\delta_{(p\sigma)\, {\not\in}\, A} , \nonumber \\
{a}_4(p) &=& \sum_{q}{a}_0 ({q\downarrow}) {a}_0
({p-q\uparrow}), \nonumber
\end{eqnarray}
where ${\bar \sigma}=-\sigma$; in block B, the corresponding 
operators can be obtained by changing $(a, A)$ into $(b, B)$. 
They are the basic operators whose matrix elements are 
kept and updated in our DMRG iterations. The total number 
of these composite operators is 6N at each sub-block.  
Using these composite operators, 
the Hamiltonian (\ref{hamiltonian2}) can be rewritten
as $H =H_{\bullet}+H_A+H_B+H_{A\bullet}+H_{B\bullet}
+H_{AB}+H_{A\bullet B}$,  
where $H_\bullet = \epsilon_k n_{k\sigma}$, 
$H_A$ is  the same as the Hamiltonian (\ref{hamiltonian2}) 
with $c_{k\sigma}$ replaced by $a_0(k\sigma )$, 
\begin{eqnarray}  
H_{A\bullet}&=& {U\over N}\left\{ n_{k\sigma} {a}_1(0,{\bar
\sigma})+  c^\dagger_{k\sigma} {a}_3(k\sigma
) +  {a}_3^\dagger (k\sigma) c_{k\sigma}\right\} ,\nonumber \\ 
H_{AB}&=& {U\over N}\sum_{p} \Bigl\{ \sum_{\sigma^\prime} 
\Bigl[ {1\over 4} {a}_1(p\sigma^\prime ) {b}_1(-p
{\bar\sigma}^\prime )+ {b}_0^\dagger (p\sigma^\prime
){a}_3(p\sigma^\prime ) \nonumber \\ 
&&+ (a\longleftrightarrow b) \Bigr] 
+{a}_4^\dagger (p){b}_4(p)-{b}_2(p)
{a}_2^\dagger (p)\Bigr\} +h.c. ,\nonumber\\
H_{A\bullet B}&=&{U\over N} c^\dagger_{k\sigma}
\sum_{p}\Bigl\{ {a}_1(k-p{\bar\sigma} ){b}_0(p\sigma )
-{\tilde a}_2 (p-k){b}_0(p{\bar\sigma})
 \nonumber\\
&& +\sigma {b}_0^\dagger (p{\bar\sigma} ){a}_4(k+p)
)+(a\longleftrightarrow b)\Bigr\} +h.c.,\nonumber
\end{eqnarray} 
and $H_B$ and $H_{B\bullet}$ can be obtained from $H_A$ and
$H_{A\bullet}$ by changing $a$ into $b$.  In $H_{A\bullet
B}$, ${\tilde a}_2 (p)= {a}_2^\dagger (p)$ if $\sigma
=\uparrow$ or ${a}_2 (p)$ if $\sigma =\downarrow$.

The breaking of lattice symmetries in truncating basis
states is a unavoidable source of errors in the DMRG method.
It is also the main difficulty in using the DMRG
method to in real space in 2D.  
In real space, the translation symmetry, i.e.  the momentum
conservation, is broken in the DMRG method.  In momentum
space, however, this symmetry is preserved.  This symmetry
property is undoubtedly very useful.  Combined with two other
conserved quantities, the number of up spins  
$N_\uparrow$ and the number of down spins  
$N_\downarrow$, it can be used to block diagonalize the
Hamiltonian.  Not only can this save the computer time, but
also it allows us to keep many more eigenstates in the
truncation of Hilbert space as the number of non-zero matrix
elements is now significantly reduced.  Basis states at each
block can be classified by three quantum numbers $(N_\uparrow
, N_\downarrow , P)$ with $P$ the total momentum.  For the
composite operators defined in (\ref{composite}) it can be
shown that their matrix elements are nonzero only
when the difference between the momentum of the initial
state and that of the final state is $p$.

In the DMRG method, the basis states for both the system
block $A$ and the environment block $B$ are incomplete.  If
the Hamiltonian contains terms with interactions between $A$
and $B$, the matrix elements of these terms will be less
accurately approximated after the truncation of the Hilbert
space compared with the matrix elements of other terms which
are defined purely within each block.  In 1D in real space
one can choose $A$ and $B$ so that no interactions exist
between $A$ and $B$, i.e.  $H_{AB}=H_{A\bullet B}=0$, if all
terms in the Hamiltonian are local in space.  In real space
in 2D or in momentum space in any dimension, however,
interaction terms between $A$ and $B$ always exist no matter
how $A$ and $B$ are constructed.  Thus in general results
obtained by the DMRG method in real space in 2D or in
momentum space in any dimension will not be as accurate as,
for example, the ground state energy of the 1D spin-1
Heisenberg model that White and Huse obtained using the
real-space DMRG method\cite{heisenberg}.

In momentum space, different size lattices have different
$k$ points.  Thus in using the DMRG method in momentum
space, the size of the lattice needs to be fixed at the
beginning.  This means that only the finite-size approach of
the DMRG method can be used in momentum space.  To use the
finite-size approach, however, one needs first to build up a
series of initial system blocks and the corresponding
environment blocks.  As the infinite lattice approach of the
DMRG is not applicable in this case, we shall use the
conventional RG method\cite{tao,white2} to build up the
initial system and environment blocks.  The steps in
building up the initial system blocks (similarly for the
environment blocks) are as follows:  1.  Start from a small
system block $A_1$, which can be handled without truncation
of basis states.  2.  Add a new ($k\sigma$) point to $A_1$
to form a new system block $A_2$.  3.  Diagonalize the
Hamiltonian in the Hilbert space spaned by $A_2$.  4.
Truncate the Hilbert space by retaining $m$ lowest energy
eigenstates, but restrict the number of states retained at
each $(N_\uparrow , N_\downarrow , P)$ subspace not more
than a small integer $n$.  Here $n$ is a variational
parameter which should be determined so that the final
result for the ground state energy is minimized.  In our
calculations, we find that $n=1$ or 2 generally gives the
best result for the ground state energy for $8\times 8$
systems with 1000 states  retained.  
The reason for limiting the number of retained
states at each $(N_\uparrow , N_\downarrow , P)$ subspace is
to prevent the retained states from being centralized in a few
$(N_\uparrow , N_\downarrow , P)$ subspaces in $A_2$.
Otherwise, some of the $(N_\uparrow , N_\downarrow , P)$
subspaces in $A_2$, which may have substantial contribution
to the final ground state energy, may be neglected in the
DMRG iterations later.  5.  Replace $A_2$ by $A_1$ and
repeat the steps 2-5 until all the initial system blocks
required are established.

All $(N_\uparrow, N_\downarrow , P)$ 
subspaces that a system block  can have are determined purely 
by the ($k\sigma$) points the block contains. 
In the above initialization step, in fact not all 
$(N_\uparrow, N_\downarrow , P)$ subspaces in a
system (or environment) block need  be considered.  
If a subspace with the quantum numbers 
$(N_{1,\uparrow}, N_{1,\downarrow} , P_{1})$  in a 
system (environment) block  
can not find a subspace with the quantum numbers
$(N_{2,\uparrow}, N_{2,\downarrow} , P_{2})$ in the 
corresponding environment (system) block such that 
$(N_{1,\uparrow} +N_{2,\uparrow}, N_{1,\downarrow} 
+N_{2,\downarrow}, P_{1}+ P_{2})$ is equal to the 
quantum numbers of the ground state required, then 
this $(N_{1,\uparrow}, N_{1,\downarrow} , P_{1})$ subspace 
in the system (environment) block will have no 
contribution to the ground state and can be  ignored.

After the above initialization step, the {\it finite-size
approach of the DMRG method} will be used to find out
the eigenvalue and eigenfunction of the ground state for a given
filling factor and momentum.  Now $m$ largest eigenstates of
the {\it density matrix} are retained at each truncation of
Hilbert space and the number of retained states at each
$(N_\uparrow , N_\downarrow , P)$ subspace will no longer be 
limited.

We have evaluated ground state energies of the Hubbard model
in both 1D and 2D using this momentum-space DMRG method. 
In most of our calculations 1000 states are retained at 
each truncation of basis states. The number of states 
retained at each $(N_\uparrow , N_\downarrow , P)$ subspace 
on average is small. 
Our results are much better than those obtained by the
conventional RG method\cite{white2}.  For a 10-electron
system with U=4 (the energy is measured in unit of 
t, i.e.  t=1) in 2D, 
for example, the momentum-space
DMRG result for the ground state energy with 1000 states
kept is $-19.57$, which is much lower than the conventional
RG result with even 3000 states kept, $-18.541$
\cite{white2} (their relative errors with respect to the
exact diagonalization result\cite{parola} are 0.05\% and
5\%, respectively).

Table I compares the ground state energies obtained by the
DMRG method in momentum space with those obtained in real
space\cite{wang} at half-filling in 1D.  The ground state
energies obtained by the DMRG method are lower in real space
than in momentum space if the same number of states is kept,
which means that in 1D the DMRG method works better for the
Hubbard model in real space than in momentum space.
This is not surprising because the Hubbard interaction is a
local interaction and at half filling all electrons are localized in
space as a result of a 
Mott insulator transition. 

Table I\/I compares the momentum-space DMRG results with the
exact diagonalization and the quantum Monte-Carlo results on
2D square lattices.  The largest lattice we have studied 
so far is 12$\times$12. Compared with the exact results on 
4$\times$4 lattice, we
find that the DMRG results are very accurate when $U$ is
small.  The relative error for the DMRG result with $U=2$ is
$3\times 10^{-4}$.  The momentum space DMRG method works
better in the weak coupling limit because the single
particle basis state used here is the plane wave state.
When U=0, the ground state is a filled Fermi sea of
non-interacting electrons, the momentum-space DMRG method
gives the exact result for the ground state even when only
one state is kept.  For large $U$, the DMRG results are 
not as good as in the weak coupling limit.  
However, they are still comparable with
those obtained by the projected quantum Monte Carlo and the
stochastic diagonalization methods.  Both the DMRG method
and the stochastic diagonalization method satisfy the
variational principle (the projected quantum Monte Carlo
method does not satisfy the variational principle due to 
the important sampling), and the results obtained from these two
methods give upper bounds of ground state energies.  On
a $8\times 8$ lattice, the ground state
energies obtained by the DMRG method are systematically
lower (hence better) than the stochastic diagonalization
results.  The DMRG results therefore provide new upper bound
for the ground state energies of the Hubbard model  on these
lattices.

In 1D, no exact  diagonalization  results are available  for
the 16-site Hubbard model, but the results obtained with the
real space DMRG method are very  accurate\cite{wang}.  If we
use the best result for the ground state energy of the 1D 16
site Hubbard model obtained by the real-space DMRG method as
the true value of the ground state energy, we estimate  that
the relative  error for the ground state energy  obtained by
the momentum-space  DMRG method with m=1000 is 5\% when U=4,
which is higher  than the  corresponding  value in 2D.  Thus
the momentum  space DMRG method works better for the Hubbard
model  in 2D than in 1D.  Physically  this  is  because  the
contribution of the kinetic energy term to the ground state,
which  is  rigorously  treated  in the  momentum-space  DMRG
method, is larger in 2D than in 1D.

The above discussion for the momentum-space DMRG method has
been focused mainly on the Hubbard model.  However, it can
be easily generalized to apply to several other physically
interesting models, such as the Anderson lattice model and
the interacting fermion model with nearest-neighbor Coulomb
potentials. For the Anderson lattice model, the composite
operators defined in (\ref{composite}) can be used without
modification.  In other cases we need to generalize the
definitions of the composite operators in (\ref{composite})
(for example $a_3$ in (\ref{composite}) should be defined as
$a_3(p\sigma ) = \sum_{q_1q_2\sigma^\prime}V(q_1-q_2)
{a}^{\dagger}_0 ({q_1 \sigma^\prime}) {a}_0
({q_2\sigma^\prime}) {a}_0 ({p+q_1-q_2\sigma})
\delta_{(p\sigma)\, {\not\in}\, A} $ for a general
electron-electron interacting model $\sum V(q-q^\prime )
c^\dagger_{q\sigma} c_{q^\prime\sigma}
c^\dagger_{q^{\prime\prime} +q^\prime \sigma^\prime}
c_{q^{\prime\prime}+q\sigma^\prime} $) and introduce some
new composite operators of electrons.  For an arbitrary
interacting fermion model with a Coulomb-type potential, the
number of composite operators required is generally of order
$N^2$, which will limit the application of the
momentum-space DMRG method to small lattices.  However, if
$V(q-q^\prime)$ can be factorized as a sum of products of a
function of $q$ and a function of $q^\prime$, i.e.  $V(q-
q^\prime )= \sum_l^n f_l(q) g_l (q^\prime )$, it can be
shown that the total number of composite operators required
can be reduced to the order of the system size $N$.  In that
case a 12$\times$12 or even larger lattice system is
accessible by the momentum-space DMRG method with presently
available computer facilities.  The Hubbard model is
obviously a factorizable potential, with $f_1=g_1=1$ and
$n=1$.  The nearest-neighbor Coulomb potential is also a
factorizable potential:  $V(q-q^\prime ) \sim \cos
(q-q^\prime ) = \cos q \cos q^\prime + \sin q \sin
q^\prime$.  For this nearest-neighbor Coulomb potential, the
total number of composite operators needed is 5$N$ in 1D and
7$N$ in 2D if fermions are spinless.

In conclusion, we have generalized successfully 
the DMRG method to momentum space 
and studied ground state properties of the Hubbard
model.  Our results show that the momentum-space DMRG method is 
a powerful numerical method for studying the Hubbard 
model in 2D. There is no reason to believe that this is 
the only model for which the momentum-space DMRG method can work. 
For any finite ranged potential, if it is factorizable,  
we believe that the momentum-space DMRG should 
work even better than for the Hubbard model.

The numerical calculations reported here were performed on a
HP735/99 workstation.  In the DMRG iteration, all
intermediate data were stored in a hard disc.
For the calculation on an $8\times 8$ (12$\times$12) 
lattice with 1000 states kept the computer memory 
space needed is 36 (45) Mb in RAM and
200 (500) Mb  in hard disc.  

We would like to thank G.  A.  Gehring and R.  J.  Bursill
for fruitful discussions and Xiaoqun Wang for supplying the
real-space DMRG results in Table I.

\begin{table}

\caption{Comparison of the ground state energy per site
obtained using the DMRG method in momentum space with that
obtained in real space at half-filling on a 16-site chain.}

\begin{tabular}{ccccc}

& \multicolumn{2}{c}{Momentum space} &
\multicolumn{2}{c}{Real space\cite{wang}} \\

m & U=1 & U=4 & U=1 & U=4 \\ \hline

400 & -1.02925 & -0.51316 & -1.02958 & -0.575896 \\

600 & -1.02944 & -0.53574 & -1.02969 & -0.575900 \\

800 & -1.02952 & -0.53724 & -1.02972 & -0.575901 \\

1000 & -1.02958 & -0.54562 & & \\ 1200 & -1.02959 & -0.55218
& &

\end{tabular}

\vspace{10mm}

\caption{Comparison of the ground state energy obtained
using the DMRG method with m=1000 with the
exact results on 4$\times$4 lattices (Exact), 
the cluster diagonalization 
results on 6$\times$6 lattices (CD), 
the projected quantum Monte Carlo
(QMC) and the stochastic diagonalization
(SQ) results on 2D square lattices with N electrons.}

\begin{tabular}{ccccccc}

$L_x\times L_y$ & U & N &
Exact\cite{parola}/CD\cite{riera} & DMRG & QMC\cite{raedt} 
& SQ\cite{raedt}
\\ \hline

4$\times$4 & 2 & 16 & -18.01757 
& -18.012  &  &  \\

4$\times$4 & 4 & 14 & -15.74459 & -15.673 & & \\ 

4$\times$4 & 4 & 16 & -13.62185
& -13.571  & -13.6 & -13.59 \\ 

4$\times$4 & 8 & 16 & -8.46887  
& -8.263   &  -8.48 & \\

6$\times$6 & 4 & 26 & 41.49 & 41.108 & 41.98 & 40.77 \\

8$\times$8 & 4 & 10 &  
& -34.325 & -34.3 & -34.31 \\

8$\times$8 & 4 & 18 &  
& -54.394 & -54.6 & -54.37 \\

8$\times$8 & 4 & 26 & 
& -66.098 &  -66.8 & -66.05 \\

12$\times$12 & 4 & 18 & & -64.107 & & 

\end{tabular}

\end{table}

\end{document}